# Modeling of Λ graded In(x)Ga(1-x)N solar cells: comparison of strained and relaxed features

**Mirsaeid Sarollahi[a], Mohammad Zamani Alavijeh[e], Manal A. Aldawsari[b], Rohith Allaparthi[a], Reem Alhelais[b], Malak A. Refaei[b], Md Helal Uddin Maruf [c], Morgan E. Ware[a,b,c,d*]**

[a]University of Arkansas, Electrical Engineering Department, 3217 Bell Engineering Center, Fayetteville, AR 72701
[b]University of Arkansas, Microelectronics-Photonics Program, 731 West Dickson Street, Fayetteville, Arkansas 72701,
[c]University of Arkansas, Material Science and Engineering 731 West Dickson Street, Fayetteville, Arkansas 72701,
[d]Institute for Nanoscience and Engineering, Fayetteville, Arkansas 72701
[e]University of Arkansas, Physics Department, Fayetteville, AR 72701

**Abstract**. The optical properties of Λ graded InGaN solar cells are studied. Graded InGaN well structures with the indium composition increasing then decreasing in a Λ shaped pattern have been designed. Through polarization doping, this naturally creates alternating p-type and n-type regions. Separate structures are designed by varying the indium alloy profile from GaN to maximum indium concentrations ranging from 20% to 80%, while maintaining a constant overall structure thicknesses of 100 nm. The solar cell parameters under fully strained and relaxed conditions are considered. The results show that a maximum efficiency of $\cong$ 5.5%, under fully strained condition occurs at x=60%. Solar cell efficiency under relaxed conditions increases to a maximum of 8.3% at 90%. While Vegard's law predicts the bandgap under relaxed conditions, a Vegard-like law is empirically determined from the output of Nextnano for varying In compositions in order to calculate solar cell parameters under strain.

**\*Corresponding Author**, E-mail: meware@uark.edu

## 1 Introduction

Ternary alloys of Group III-N materials are great candidates to be used in photovoltaic devices. This is due to interesting properties such as high thermal conductivity, high optical absorption, and high radiation resistance[1–4]. In addition, the direct band gap, which is tunable over most of the usable solar spectrum[5,6] makes them an appropriate choice for photovoltaic devices[7–9]. Several simulation studies for InGaN homojunctions have been reported. The solar efficiency, $\eta$, of a single junction $In_{0.65}Ga_{0.35}N$ solar cell was reported to be 20.28%[10], for example. However, a higher efficiency of $\eta$ =24.95% was demonstrated using the same indium composition due to adoption of the density of states (DOS) model[11]. This presented much more information about recombination/generation in the solar cell than the lifetime model by neglecting defects.

Additionally, similar work for single-junction $In_{0.622}Ga_{0.378}N$ demonstrated an efficiency of, $\eta =26.5\%$, when the optical properties and physical models such as the Fermi-Dirac statistics, Auger and Shockley-Read-Hall recombinations, and the doping and temperature-dependent mobility model were taken into account in the simulations[12]. In another example[9,11–13] InGaN homojunction solar cells were shown to have parameters for the short circuit current, open circuit voltage, fill factor, and efficiency equal to ~31.8 mA/cm$^2$, ~0.874 V, ~0.775, and 21.5%, respectively, which were determined after varying the thicknesses of the junctions[13].

Growing p-type doped InGaN alloys are still considered challenging and problematic[14–17], therefore polarization doping plays an important role in the designed Λ shape graded structure. This grading of the film in group III-nitride materials can achieve very high levels of doping, without using additional impurities in the lattice[18]. This can be understood as follows. InN and GaN contain different spontaneous crystal polarizations (P). When a graded structure is grown, polarization doping is introduced correlating to the spatial rate of composition change by $-\nabla.P = \rho$. When the indium composition is increasing from GaN to InGaN, the background charges are negative ($\rho < 0$) which attract positive charges (holes) to make p-type doping. On the hand, when indium composition is decreasing from InGaN to GaN, the background doping charges are positive ($\rho > 0$) which attract negative charges (electrons) to make n-type doping[19]. In contrast, other reports show that AlGaN polarization doping due to graded films are reversed in comparison with InGaN. That is, by increasing the aluminum composition from GaN to AlGaN background charges are positive which create n-type doping due to attracting electrons, and vice versa[20]. These polarization effects include both spontaneous and piezoelectric polarizations in III-V nitride heterostructures[21]. Additionally, advantage of using graded structure is that highly strained materials can be obtained with reduced dislocation and phase separation whereas growing a thick layer with a fix Indium composition would be challenging due to high relaxation.

In this work, we present the simulation of a Λ-shaped graded structure. This is comprised of a double graded structure of $In_xGa_{1-x}N$ in which the indium composition starts to increase linearly from GaN to $x_{max}$, then linearly decrease from $x_{max}$ to GaN. As explained above, this results in a p-n junction. The whole thickness of the structure is 100 nm (chosen to compare with previously published data)[22,23], and $x_{max}$ varies in different structures from 20% to 90%. The indium composition, strain relaxed energy band diagram, and fully strained energy band diagram versus position for x=50% are shown in Fig. 1 a, b & c respectively. According to Fig.1b and c, it is clear that the Λ graded structure creates p n junction. The first layer (0 – 50 nm) creates p-type (GaN to InGaN) and the second layer (50-100 nm) creates n-type (InGaN to GaN). (-10 to 0 and 100 to 110) are defined as ohmic contacts at the two edges and have no intrinsic composition.

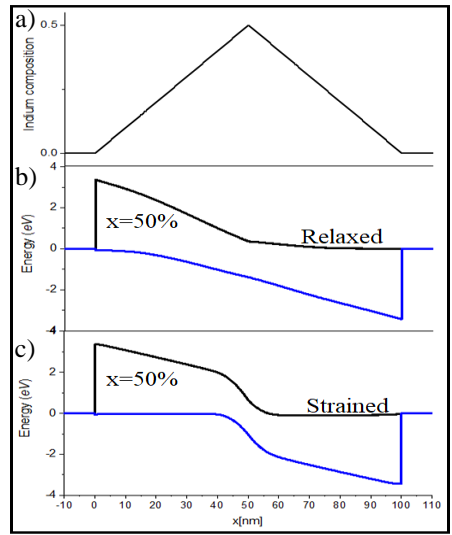

**Fig.1.** a) Indium composition vs position, b) Energy band diagram vs position (fully relaxed), c) Energy band diagram vs position (fully strained) for Λ graded structure at x=50%

## 2 Simulation

Nextnano software, which has two varieties, nextnano3 and nextnano+ with slightly different specialities, was used to simulate the solar cell parameters. The software is able to calculate the generation rate for a constant alloy only. So, our graded structure is approximated as a step graded structure. Then, the generation rate of each layer is calculated in nextnano3. These are compiled manually into the generation rate versus position for the whole graded structure and subsequently imported in nextnano+ to calculate the light IV curve, incorporating the entire continuous graded film. This is then used for calculating solar cell parameters such as short circuit current, open circuit voltage, fill factor and efficiency for each structure.

The Λ-graded structure is defined by a *d* nm layer from GaN to $In_{x_{max}}Ga_{1-x_{max}}N$ followed by another d nm layer grading back to GaN. The total thickness of the designed structure is then *2d*. To obtain the generation rate for the graded structure we need a method to convert it to some set of constant alloy layers. A simple formula is created for this. A step size provides the thickness of each constant composition layer, such that the Indium composition at each position increases by:

$$\frac{x_{max}}{d-(step\ size)} \tag{1}$$

We used a step size 1 nm (50 layers on each side of the maximum for a total of 100 layers for our step graded structure). This step graded approximation of our structure is shown schematically in Fig. 2.

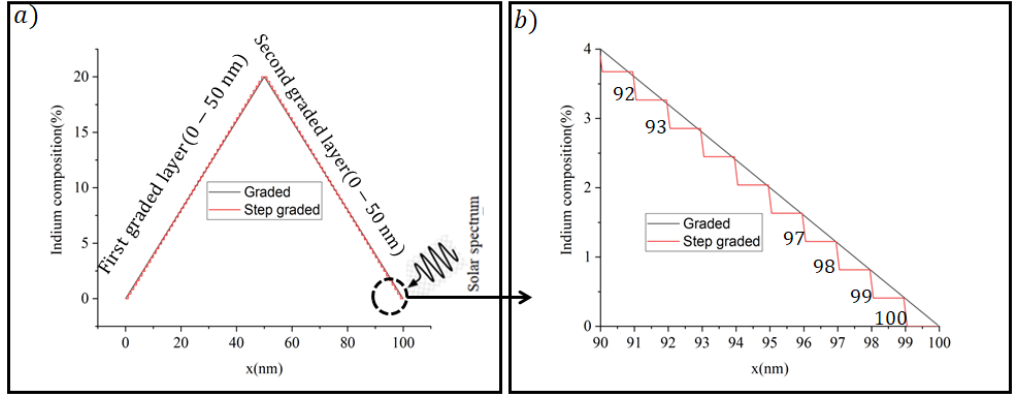

**Fig. 2** a) Converting graded structure to step graded by equation 1 to calculate generation rate in nextnano. Receptivity of layers is given by the number below each layer between (90-100 nm) when light passing through the structure. Layers numbers are written next to each layer. b) Zoom in for 90-100 nm to show converting graded to step graded

The generation rate for each layer in the step graded structure should be calculated in order to find the generation rate for the whole structure. To find the generation rate in each layer, the absorption coefficient and the light intensity should get imported into nextnano3. The solar spectrum with intensity, $I_0$, first enters the structure through layer 100 (1 nm GaN), then the imported $I_0$ along with the GaN absorption coefficient ($\alpha$) are used by nextnano3 to calculate the generation rate within this layer, for the first nm of penetration. The light intensity decreases as it passes through the layers. For layer 99, $I_{100} = I_0 \exp(-\alpha_{100} d)$, where $d$ is the thickness of layer 100 and $\alpha_{100}$ is the absorption coefficient of that layer. By continuing this method for all given layers, the generation rate versus position is obtained as $G_1(0-1)$, $G_2(1-2)$,… $G_{98}(97-98)$, $G_{99}(98-99)$, $G_{100}(99-100)$. Then, G vs position is imported into nextnano+ in order to determine the illuminated *I-V* curve for the solar cells. Other solar cell parameters can then be extracted from the light *I-V* curve.

## 3    Effects of strain on optical properties

There is a large lattice mismatch of 11% between InN and GaN. Due to this, strain is created for layers that are grown on a GaN substrate and can impact a device in different ways. First, the position of the valence and conduction band edges change, which reshapes the potential well along with the confinement of electrons and holes in the active region. The strain also changes the effective mass of carriers and the density of states[24–28]. At the same time, general parameters which are derived from the bandgap. like the absorption coefficient are modified. Additionally, the k.p Hamiltonian of the Schrödinger equation is modified. So a detailed understanding of the strain is an important parameter for design, development and study of electronic and optical properties of semiconductor heterostructures based devices[29].

Nextnano software considers strain and, both piezoelectric polarization and spontaneous polarization[30]. Here, we assume InGaN to be strained to a GaN substrate.

For modeling the absorption coefficient of InGaN as the alloy varies linearly, the bandgap under strain is required. In the absence of strain (fully relaxed) the $In_xGa_{1-x}N$ bandgap is calculated using Vegard's law which is defined in equation(3):

$$E_g(In_xGa_{1-x}N) = E_g(InN)x + E_g(GaN)(1-x) - b(x)(1-x), \qquad (3)$$

Large incosistencies are reported in literature for the bowing parameter, *b*, in the range of 1.4 to 3 eV which is generally thought to be the result of unaccounted for strain states in the $In_xGa_{1-x}N$ layers studied[31–37]. Additionally, there is some debate about whether the bowing coefficient parameter is compositionally dependent[38,39] or not[40–42]. We derive our values from the nextnano materials database as a good overal average[43–45].

The values for the bandgap energies under strain and relaxed as functions of composition, *x*, were determined using nextnano. The output was fit to a second order polynomial in *x*, similar to Vegard's law in order to achieve an analytic representation for this quantity[46]. Both are shown Eq. 4 based on Fig. 3.

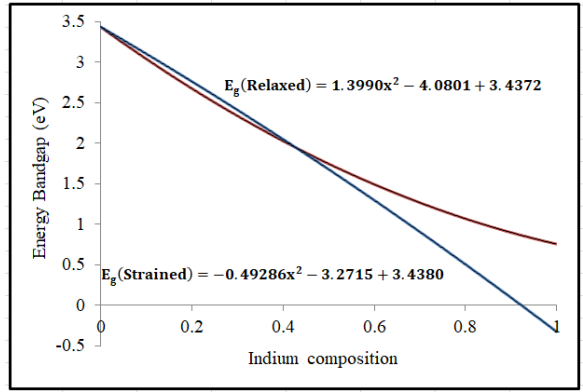

**Fig. 3** Energy bandgap Vs Indium Composition under relaxed (Red line) and strained (Blue line)

$$E_{g(Relaxed)} = 1.3990x^2 - 4.0801x + 3.4372 \qquad \text{4a)}$$

$$E_{g(Strained)} = -0.49286x^2 - 3.2715x + 3.4380 \qquad \text{4b)}$$

According to equation 4b), the bandgap for strained $In_xGa_{1-x}N$ with $x>~90\%$ is negative[46] implying that InGaN is a semi-metal in this range[47,48]. This simulation output about negative bandgap for high indium compositions get confirmed with new report which indicates polarization field in strained InN quantum well inverts conduction band and valence band which is precondition for topological insulator[49]. In this work, we do not further discuss this subject, but materials with negative bandgap potentially present novel research directions. Here, all simulations are performed up to 90%. For the relaxed material, however, there is no problem with the bandgap becoming negative and in fact relaxed devices with $x_{max}>90\%$ have been simulated and will be reported on separately as they demonstrate separate and interesting anomalous response. The limitation of this work is the possibility of having fully strained and fully relaxed structures as a real device.

The absorption coefficient for InGaN ternary alloys can be parameterized as follows[50–52]:

$$\alpha(\lambda) = \alpha_0 \sqrt{a(x)(E - E_g) + b(x)(E - E_g)^2} \qquad (5)$$

with $E$ the photon energy, $E_g$ the bandgap, and $\alpha_0 = 10^5 \, cm^{-1}$. In addition, $a(x)$ and $b(x)$ are dimensionless parameters which are obtained by fitting over the entire composition range[50].

$$a(x) = 12.87x^4 - 37.79x^3 + 40.43x^2 - 18.35x + 3.52 \quad (6)$$

$$b(x) = -2.92x^2 + 4.05x - 0.66 \quad (7)$$

Above parameters were obtained from Ref.47.

As a result, the absorption coefficients as functions of bandgap can be calculated empirically over the entire range of alloys. Once this calculation has been perform Eq.5 can be converted to functions of the bandgap energies under strain by plotting and fitting the parameters, $a$ and $b$, as functions of the strained bandgap energy, $a(E_{g(Strain)})$ and $b(E_{g(Strain)})$. These are shown in Eqs. 8 and 9 and Fig. 4.

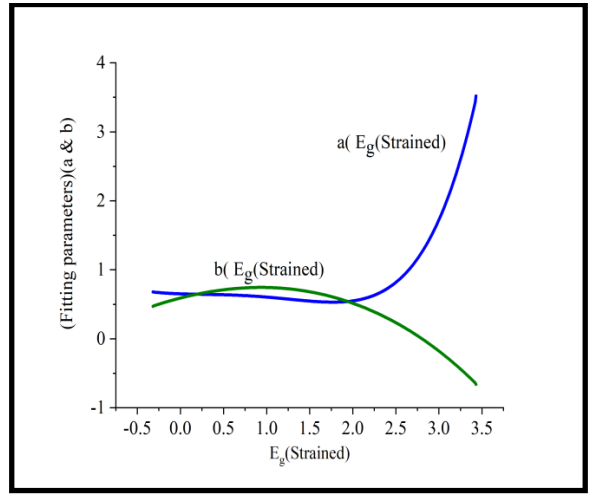

**Fig. 4** Fitting parameters (a & b) Vs Energy bandgap under strained

$$a(E_{g(Strained)}) = 1.2550 \cdot 10^{-2} E_{g(s)}{}^5 - 8.3347 \cdot 10^{-3} E_{g(s)}{}^4 - 5.8864e. \, 10^{-2} E_{g(s)}{}^3 -$$

$$4.5488. \, 10^{-2} E_{g(s)}{}^2 - 4.1689. \, 10^{-2} E_{g(s)} + 0.65425 \quad (8)$$

$$b(E_{g(Strained)}) = -1.5280 \cdot 10^{-3} E_{g(s)}^4 - 5.5727 \cdot 10^{-3} E_{g(s)}^3 - 1.6201 \cdot 10^{-1} E_{g(s)}^2 +$$
$$3.2084 \cdot 10^{-1} E_{g(s)} + 0.59135 \tag{9}$$

The absorption coefficients as functions of wavelength ($\alpha(\lambda)$) can then be calculated under strain, using the strained values for the bandgaps up to $x=90\%$. Then, using Eqs. 5 to 9, the bandgap energy and the absorption coefficient as a function of the bandgap energy can be calculated for both strained and unstrained conditions. The Vegard's law (fully relaxed) is valid only above 0.7 $eV$ (InN) so we will limit our strained study to material with bandgap above that, i.e., $x \leq 0.9$ (strained). Finally, the generation rate for each layer can be calculated after importing the absorption coefficient and light intensity relating to each layer. The generation rates for the strained condition is shown in Fig. 5. As discussed above, the generation for each nm was calculated, and then the generation for the whole structure as a function of position is compiled for both strained and relaxed.

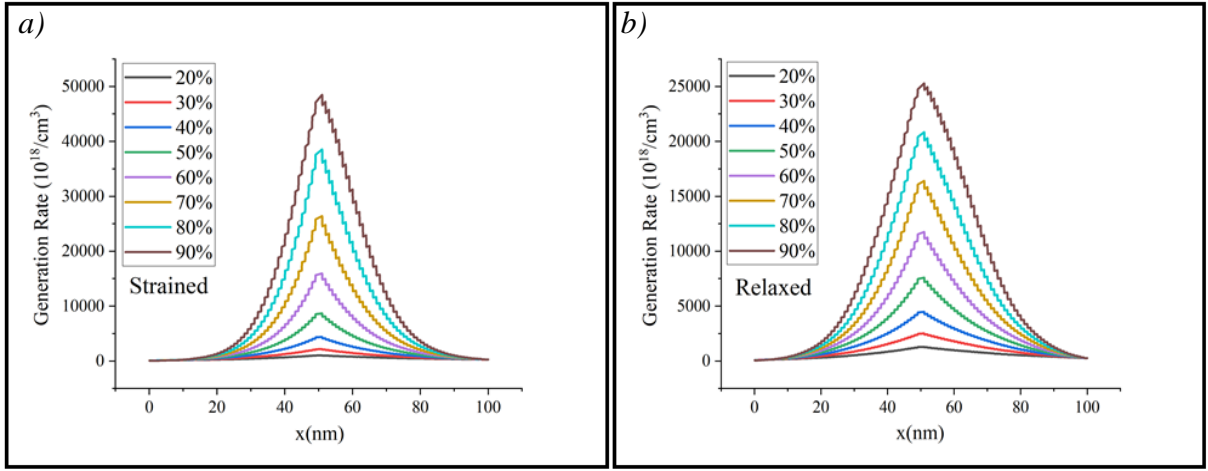

**Fig.5.** Generation rate for Λ shape graded structure under a) strained b) Relaxed condition.

## 3.1 Solar cell parameters

In order to calculate the *J-V* characterization, $AM_{1.5G}$ is used as the total irradiance on the solar cell, which supplies $100\, \frac{mW}{cm^2}$ of power to the solar cell. The *J-V* plots for the fully strained and relaxed graded structures are shown in Fig5 a) and b), respectively.

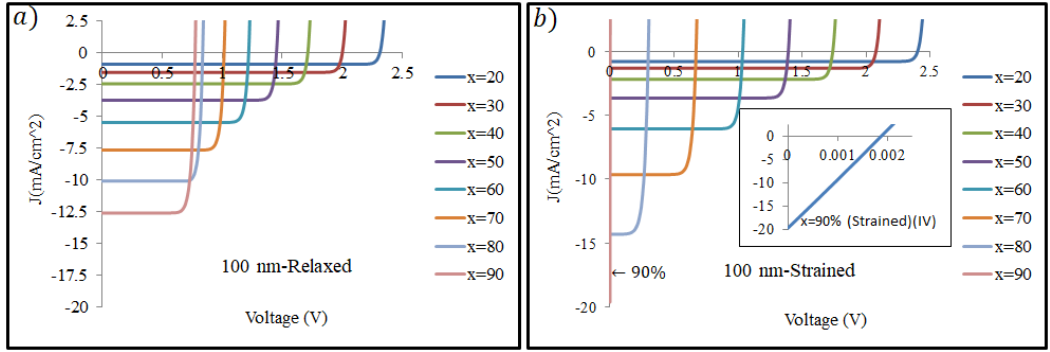

**Fig.6.** Light J-V curve under a) fully relaxed and b) strained condition for different indium compositions. Insert is J-V curve for x=90% under strained.

According to literature for single composition growth, for *x*<25% the InGaN structure should be fully strained[23] to the GaN lattice constant, however the graded films have yet to be studied to realize the critical thickness for relaxation, which may be much higher. According to Fig.7, the $V_{oc}$ effectively follows the bandgap with *x* as is also shown for different graded structures previously[23,46].

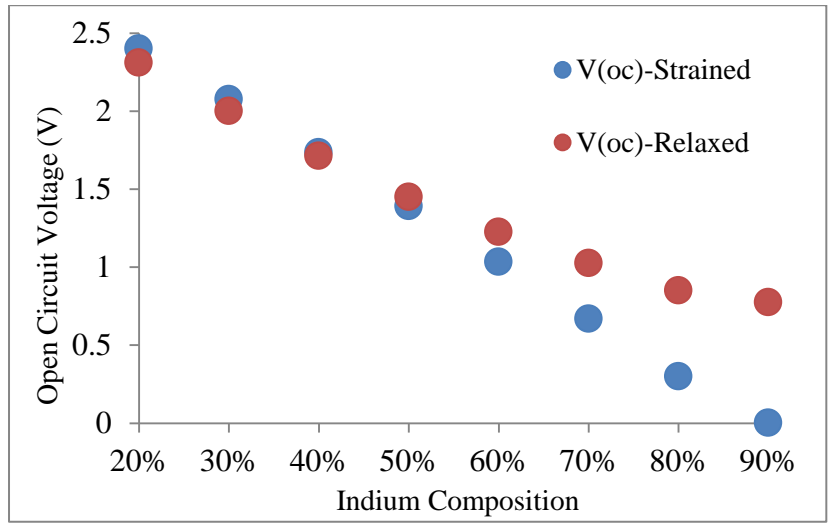

**Fig.7.** Open circuit voltage for strained and relaxed condition in different indium compositions

By increasing $x_{max}$, the $V_{oc}$ decreases due to the overall decrease in bandgap (for both strained and relaxed conditions). The short circuit current, $I_{sc}$, exhibits a reverse dependency, where increasing indium composition results in an increase in $I_{sc}$. This is shown in Fig.8. This is correct outpur according to literature. Martin A. Green [53], the maximum value of $V_{OC}$ decreases with decreasing bandgap $E_g$. This trend is reverse for the $J_{SC.}$

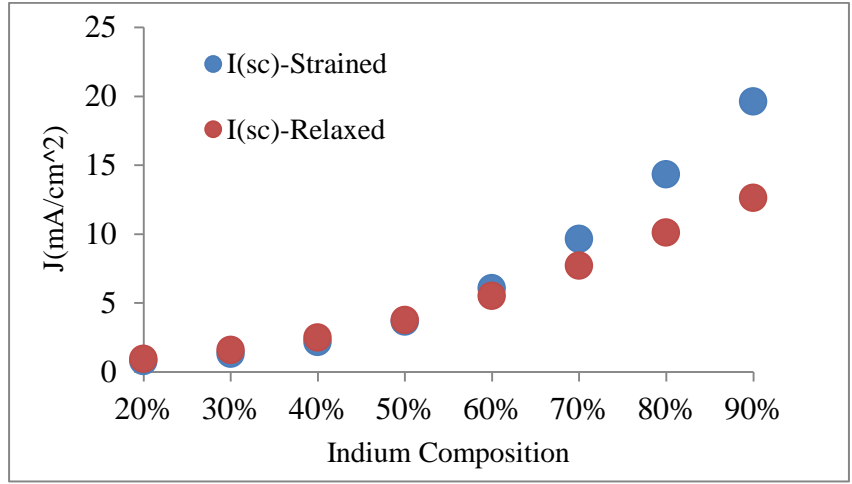

**Fig.8.** Short circuit current for strained and relaxed conditions

In addition, this output for $V_{oc}$ and $J_{sc}$ is supported by literatures. The bandgap energy of InGaN decreases by increasing Indium composition. The open circuit voltage is directly correlated to energy bandgap. In other word, The $V_{oc}$ is determined by the difference of the Fermi energies ($\Delta E_F$) of the electron and the hole in the depletion region, which in turn is affected by the bandgap energy[54]. The highest value for $V_{oc}$ is limited by bandgap energy. Lower $V_{oc}$ would be due to recombination process[53,55]. Short circuit current density $J_{sc}$ increases by decreasing the bandgap [53]. In addition, smaller bandgap (high indium) and longer penetration depth of infrared photons enhance absorption cause more generation rate (electron-hole) in high-indium-composition InGaN solar cells[56]. Low open circuit current density $J_{sc}$ in low Indium composition InGaN solar cell is the main reason of low solar efficiency in these structures. As the indium composition increasing $J_{sc}$ increases and $V_{oc}$ decreases which leads to small FF in solar cell which basically implies $V_{oc}$ is proportional to FF[54].

The fill factor (FF) is defined as the ratio between the maximum power ($P_{max} = J_m V_m$) created by the solar cell and the product, $V_{oc}J_{sc}$, as follows.

$$Fill\ Factor = \frac{J_m * V_m}{J_{sc} * V_{oc}} \quad (10)$$

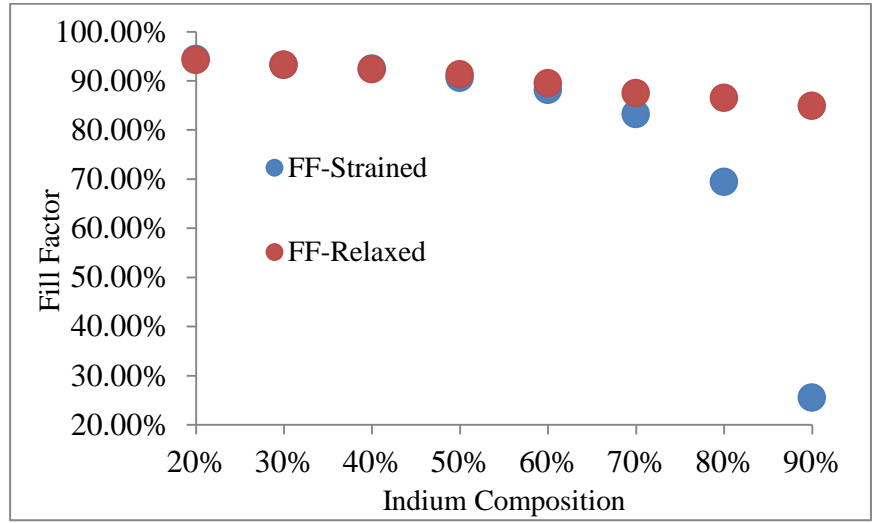

**Fig.9.** Fill Factor for strained and relaxed conditions

According to Fig.9, FF decreases as the indium composition increases. The fill factor is ~94% at $x$=20% and decreases at higher $x$, with a minimum value of ~26% at $x$=90% under strain. The minimum values is ~85% at x=90% under fully relaxed conditions. Here, the relaxed material demonstrates more stability with increasing composition than the strained material does.

Finally, the solar efficiency is reported. The solar efficiency is defined as the part of the sunlight energy that gets converted to electricity by the solar cell and is calculated by:

$$Efficiency(\%) = \frac{P_{max}}{P_{in}} = \frac{P_{max}}{100} \quad , P_{max} = FF.V_{OC}.I_{SC} \quad (11)$$

The solar efficiencies for the designed structures under fully strained and relaxed conditions are displayed for different maximum compositions in Fig.10.

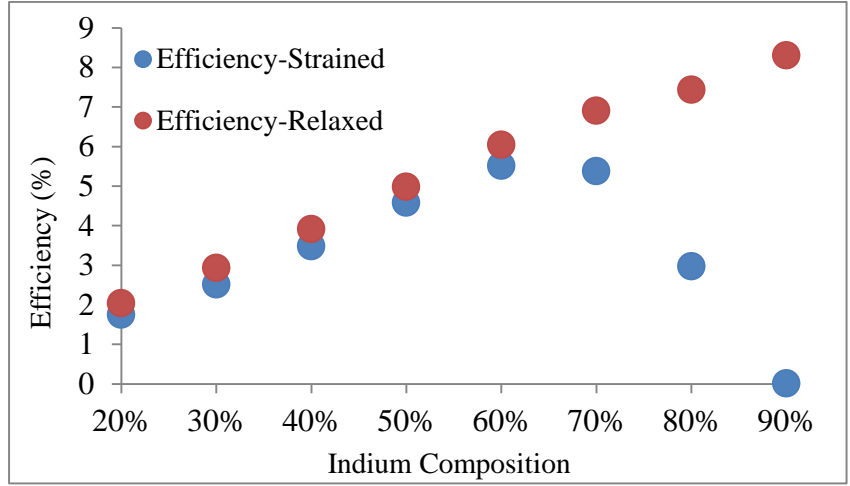

***Fig*.10.** Solar efficiency for strained and relaxed conditions

The solar efficiency for strained InGaN shows a maximum value at *x*=60% which is 5.5%. This is lower than other reports due mainly to the much thinner structure[10], however the resulting composition of the maximum is similar to published single composition results[10,42]. The minimum efficiency at 90% is less than 0.1% (strained InGaN) as both $V_{oc}$ and FF are decreased in this situation and ~2% at 20% (Relaxed InGaN). However, the solar efficiency of relaxed InGaN is ~ 8.3% for *x*= 90%, and would likely increase with further increase in $x_{max}$.

The performance of solar cell out of atmosphere was studies by using AM0. The solar cell parameters were calculated under AM0( 135.3 $mW/cm^2$) for both starined and relaxed. The data show the same trend as domestic calculation. For strain, the maximum efficiency occurs at 60%. For relax, like previous reported results for AM1.5G, the efficiency increasing by increasing Indium composition. In other word, x=90% showed the maximum efficiency. Fig. 11a) shows Standard Solar Spectra for space (AM0) and domestic (AM1.5G) use, Fig. 11b) Efficiency for strained and relaxed conditions for space using (AM0). According to Fig. 11a, the intensity of AM0 is higher than AM1.5G. The maximum efficiency for both strained and relaxed conditions at AM0 is almost the same as AM1.5G.

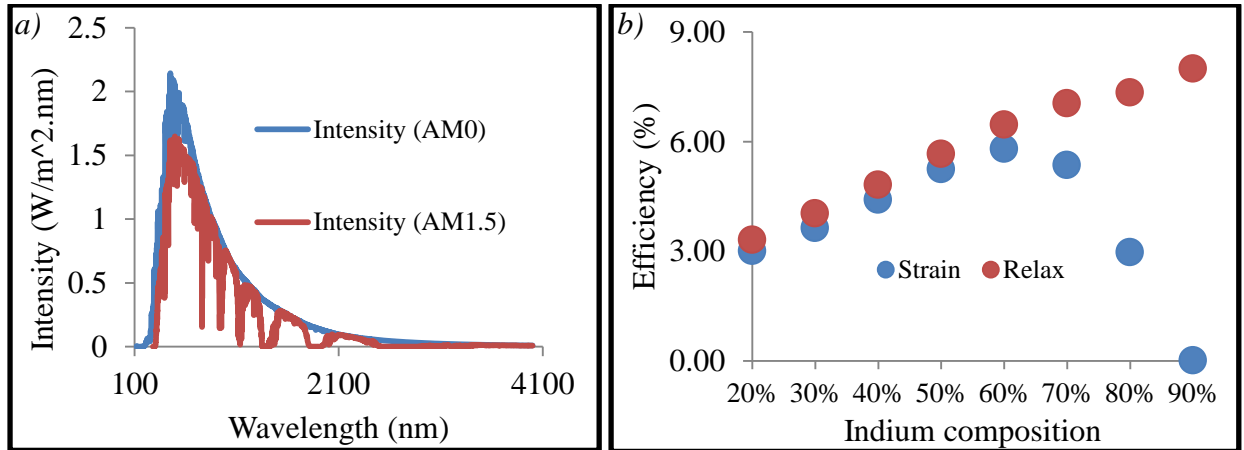

**Fig.11.** a) Standard Solar Spectra for space (AM0) and domestic (AM1.5G), b) Solar efficiency for strained and relaxed conditions under AM0

In order to investigate the effect of thickness on solar cell parameters at x=60% in strained condition, 200 and 300 nm Λ-shaped graded structures, which are defined similarly symmetric as the above structures, are studied and compared with the 100 nm structure. The *J-V* curves for the different thicknesses at *x*=60% is displayed in Fig.11.

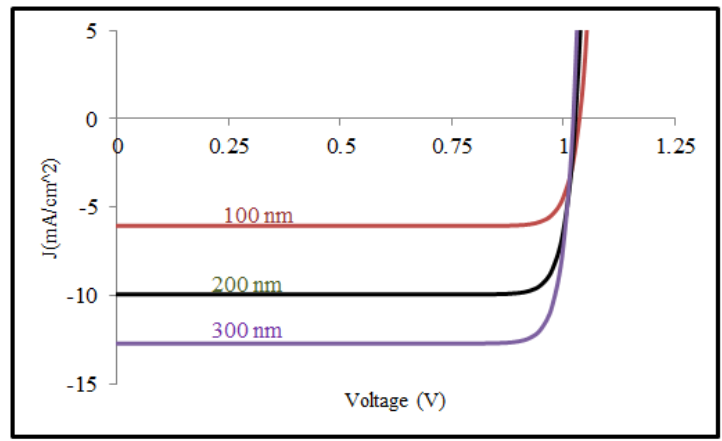

**Fig.12.** J-V curve for different thicknesses at *x*=60% for strained condition

As the solar cell thickness increases, the short circuit current increases sharply. On the other hand, the open circuit voltage decreases slowly. The solar cell parameters for different thicknesses are shown in Fig.11. With this increase in thickness, we see the following. According to Fig.12a, the $I_{sc}$ increases from $6.98\ mA/cm^2$ to $15\ mA/cm^2$. Fig.12b indicates $V_{oc}$ decreases slowly from 1.04 V to 1.026 V. Fig.12c shows the fill factor increasing from 88.12% to ~88.45%, while the overall efficiency almost doubles from 6% to 12% in Fig.12d.

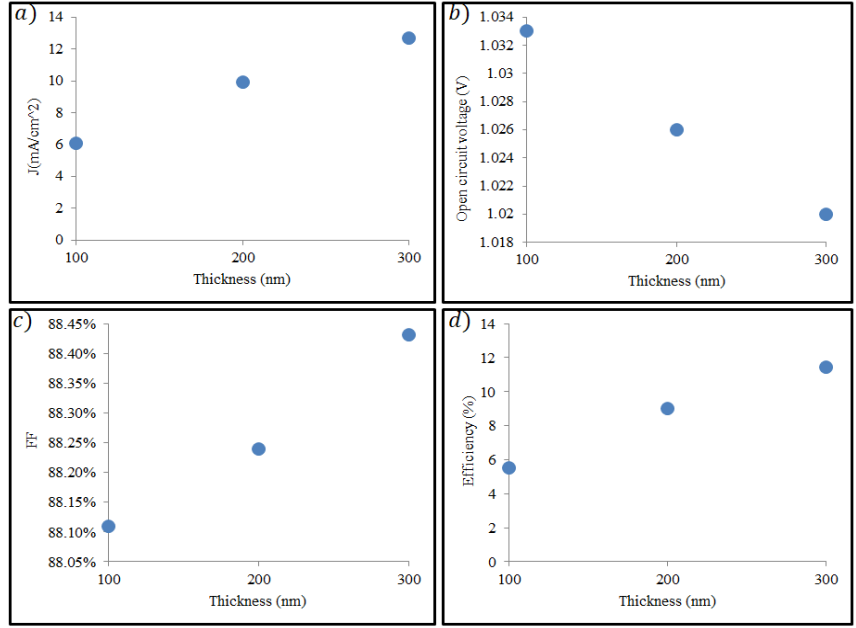

**Fig.13.** a) Short circuit current, b) Open circuit voltage, c) Efficiency (%), d) Fill Factor in different thicknesses

**Conclusion**

InGaN graded structures, where the In composition varied in a Λ-shape over only 100 nm were designed and investigated through simulation to determine their potential for use as thin film solar cell devices. The main advantage of this structure is to form p- and n- type layers due to polarization doping as p-type doping in group III-nitride materials isn't efficient. In addition, a new model (a Vegard's law-like equation) was presented to calculate the energy bandgap for InGaN strained to a GaN substrate and similarly to calculate the absorption coefficient for an arbitrary value of the InGaN bandgap. Solar cell parameters were studied for Λ-shaped structures for both strained and relaxed conditions. The maximum efficiency was found to be ~5.5% with a maximum In composition of $x_{max}(In)=60\%$ under fully strained conditions. However, for fully relaxed material, the maximum efficiency was found to be ~8.5% for $x_{max}(In)=90\%$, with an indication that further increase in composition would increase efficiency further. Finally, the efficiency for strained Λ-shaped structures with the total thickness increased to 300 nm and $x_{max}$(In)=60% increased by nearly a factor of two. These results indicate a useful window for InGaN Λ-graded devices to be used for solar cells. The polarization doping provides the highly doped contact regions without the addition of lifetime killing impurities, while the compositional grading naturally allows for higher In content and therefore stronger optical absorption without immediate relaxation due to lattice mismatch with GaN substrates.

**Caption List**

**Fig. 1** a) Indium composition vs position, b) Energy band diagram vs position for Λ graded structure at x=50%.

**Fig. 2** a) Converting graded structure to step graded by equation 1 to calculate generation rate in nextnano. Receptivity of layers is given by the number below each layer between (90-100 nm) when light passing through the structure. Layers numbers are written next to each layer. b) Zoom in for 90-100 nm to show converting graded to step graded

**Fig. 3** Energy bandgap Vs Indium Composition under relaxed (Red line) and strained (Blue line)

**Fig. 4** Fitting parameters (a & b) Vs Energy bandgap under strained

**Fig.5.** Generation rate for Λ shape graded structure under strained and relaxed.

**Fig.6.** Light J-V curve under a) fully relaxed and b) strained condition for different indium compositions. Insert is J-V curve for x=90% under strained.

**Fig.7.** Open circuit voltage for strained and relaxed condition in different indium compositions

**Fig.8.** Short circuit current for strained and relaxed conditions

**Fig.9.** Fill Factor for strained and relaxed conditions

**Fig.10.** Solar efficiency for strained and relaxed conditions

**Fig.11.** a) Standard Solar Spectra for space (AM0) and domestic (AM1.5G), b) Solar efficiency for strained and relaxed conditions under AM0.

**Fig.12.** J-V curve for different thicknesses at $x$=60% for strained condition

**Fig.13.** a) Short circuit current, b) Open circuit voltage, c) Efficiency (%), d) Fill Factor in different thicknesses